\documentclass[reprint,amsmath,amssymb,aps,nofootinbib, floatfix]{revtex4-2}

\usepackage{graphicx}
\usepackage{dcolumn}
\usepackage{bm}
\usepackage{booktabs}
\usepackage[usenames,dvipsnames,svgnames,table]{xcolor}
\usepackage{hyperref}
\hypersetup{
	colorlinks=true,
	linktocpage=true,
	citecolor=YellowOrange,
	linkcolor=violet,
	urlcolor=MidnightBlue!50!Aquamarine,
}

\newcommand{\orcid}[1]{\href{https://orcid.org/#1}{\includegraphics[height=0.3cm]{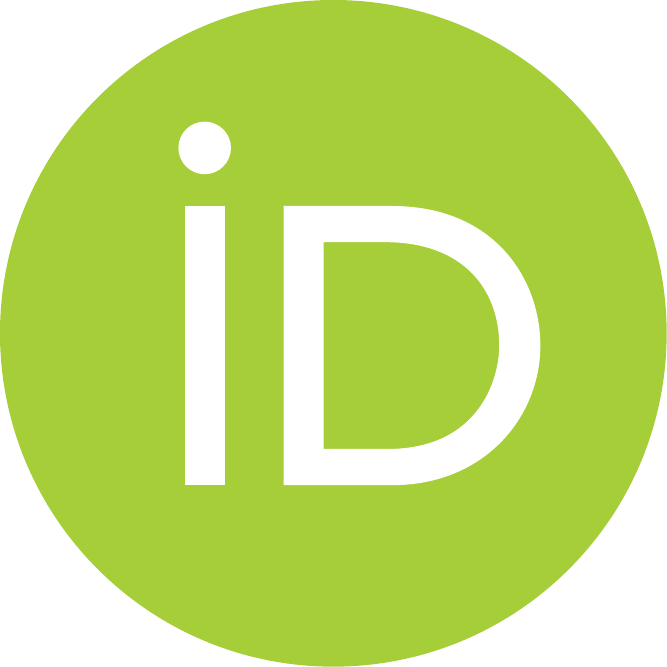}}}

\usepackage[margin=0.8in]{geometry}

\usepackage[caption=false]{subfig}
\usepackage[range-phrase=\textendash, range-units=single, separate-uncertainty=true]{siunitx}
\DeclareSIUnit\century{century}
\DeclareSIUnit\year{year}
\DeclareSIUnit\day{day}
\DeclareSIUnit\parsec{pc}

\usepackage{tikz}
\usetikzlibrary{arrows}
\usetikzlibrary{shapes}
\usetikzlibrary{shapes.geometric}
\usetikzlibrary{calc,patterns,angles,quotes}

\begin{document}

\title{Identification of neutrino bursts associated to \texorpdfstring{\\}{} supernovae with Real-time Test Statistic (RTS\texorpdfstring{$^2$}{²}) method}
\author{Mathieu Lamoureux\orcid{0000-0002-8860-5826}}
\email{mathieu.lamoureux@pd.infn.it}
\affiliation{Dipartimento di Fisica, INFN Sezione di Padova and Università di Padova, I-35131, Padova, Italy}
\date{\today}

\begin{abstract}
This paper proposes a new approach for the selection of low-energy neutrino bursts, such as the ones detected after a supernova. It exploits the temporal structure of the expected signal with respect to the more diffuse background by defining a ``Real-time Test Statistic'' (RTS) that would allow identifying very weak signals, hard to select using standard clustering methods. For a given background rate, the new method (RTS$^2$: RTS for Supernovae) increases signal efficiency while keeping the same false alarm rate for Poisson-distributed background. By adding a spatial penalty term to the definition of RTS, one can also reject spatially-correlated backgrounds such as the ones due to spallation events. Furthermore, the method is easy to implement in a real-time monitoring system as RTS can be computed recursively for successive events, and it can be easily adapted for detectors of all scales that may want to send prompt alerts e.g. through SNEWS 2.0 network.
\end{abstract}

\maketitle

\section{Introduction}
\label{sec:intro}

The detection of low-energy neutrino bursts is particularly relevant in the search for supernovae. Prompt detection would allow providing early warning for the subsequent electromagnetic observations, that is crucial to understand the underlying mechanisms.

The usual method consists in watching for upper fluctuations from the typical expected background. In most of the neutrino detectors, the main background at energies relevant for supernovae can be modelled with Poisson statistics, i.e. a constant rate with simple statistical fluctuations. For a given background rate $r$, the number of background events in a time window $w$ is simply following the Poisson distribution with $\lambda = r \times w$:
\begin{equation}
    \textrm{Poisson}(m; r,w) = e^{-r \times w} \dfrac{(r\times w)^m}{m!}.
\end{equation}

Let's denote $\tau_{\rm phys}$ the typical time scale of the physical phenomenon under study. For instance, $\tau_{\rm phys} \sim \SI{10}{\second}$ is expected in the case of supernovae. If $r \tau_{\rm phys} \gg 1$, Gaussian approximation can be used and the search for neutrino bursts can be simplified to a search of an excess with a z-score~\cite{NOvA:2020dll}.

\begin{figure}[!ht]
    \centering
    \subfloat[Sliding windows]{
    \begin{tikzpicture}[scale=0.6, every node/.style={scale=0.75}]
        \draw[-latex] (0,0)--(11, 0) node[right] {time};
        \draw[blue,<->] (0, -0.6)--(5, -0.6) node[midway, anchor=north] {$w$};
        \draw[blue,<->] (5, -0.6)--(10, -0.6) node[midway, anchor=north] {$w$};
        \draw[blue,<- ] (10, -0.6)--(11, -0.6);
        \node[blue,anchor=east] at (12.9, -0.6) {1st scan}; 
        \draw[blue,<->] (2.5, -1.1)--(7.5, -1.1) node[midway, anchor=north] {$w$};
        \draw[blue,<-] (7.5, -1.1)--(11, -1.1) node[midway, anchor=north] {$w$};
        \node[blue,anchor=east] at (12.9, -1.1) {2nd scan}; 
        \node[red] at (0.5, 0) {$\bullet$};
        \node[red] at (1.2, 0) {$\bullet$};
        \node[red] at (3, 0) {$\bullet$};
        \node[red] at (4.2, 0) {$\bullet$};
        \node[red] at (5.9, 0) {$\bullet$};
        \node[red] at (6.3, 0) {$\bullet$};
        \node[red] at (8.5, 0) {$\bullet$};
        \node[red] at (10.5, 0) {$\bullet$};
    \end{tikzpicture}
    }
    
    \subfloat[][Dynamic windows]{
    \begin{tikzpicture}[scale=0.6, every node/.style={scale=0.75}]
        \draw[-latex] (0,0)--(11, 0) node[right] {time};
        \draw[blue,<->] (0.5, -0.6)--(5.5, -0.6) node[midway, anchor=south] {$w$};
        \draw[blue,<->] (1.2, -0.75)--(6.2, -0.75);    
        \draw[blue,<->] (3.0, -0.90)--(8.0, -0.90);
        \draw[blue,<->] (4.2, -1.05)--(9.2, -1.05);
        \draw[blue,<->] (5.9, -1.20)--(10.9, -1.20);
        \draw[blue,<- ] (6.3, -1.35)--(11.0, -1.35);
        \draw[blue,<- ] (8.5, -1.50)--(11.0, -1.50);
        \node[white,anchor=east] at (12.9, -1.5) {-};
        \node[red] at (0.5, 0) {$\bullet$};
        \node[red] at (1.2, 0) {$\bullet$};
        \node[red] at (3.0, 0) {$\bullet$};
        \node[red] at (4.2, 0) {$\bullet$};
        \node[red] at (5.9, 0) {$\bullet$};
        \node[red] at (6.3, 0) {$\bullet$};
        \node[red] at (8.5, 0) {$\bullet$};
        \node[red] at (10.5, 0) {$\bullet$};
    \end{tikzpicture}
    }
    \caption{Illustration of the two standard methods to build the time windows.}
    \label{fig:cluster}
\end{figure}
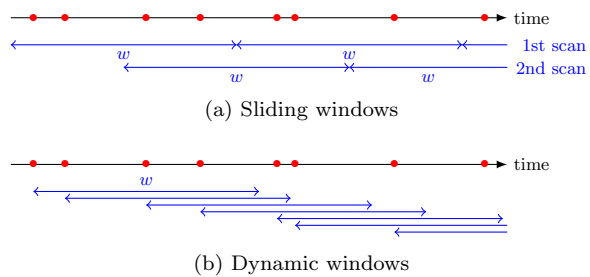

However, if the detector has a relatively low background ($r \tau_{\rm phys} \lesssim 1$), this method cannot be applied and the usual technique consists in dividing the data in clusters using time windows of size $w$, covering the physical time scale $\tau_{\rm phys}$. For supernovae, $w = \SI{20}{\second}$ is commonly used. There are two different approaches to define these time windows, as illustrated in \autoref{fig:cluster}:
\begin{itemize}
    \item Sliding windows of width $w$, with each window starting in the middle of the previous one~\cite{Agafonova:2007hn}.
    
    \item Dynamic windows of width $w$, one starting from each selected event~\cite{Abe:2016waf,Novoseltsev:2019gdt}.
\end{itemize}

It is then straightforward to define a cut on the cluster multiplicity in order to achieve a target false alarm rate $\textrm{FAR}$ (e.g. 1/year or 1/century) by using the Poisson survival function.

For sliding time windows,
\begin{equation}
    \dfrac{2}{w} \times \sum_{m=m_{\rm cut}}^{\infty} e^{-rw} \dfrac{(rw)^m}{m!} \leq \textrm{FAR},
    \label{eq:Fim_sld}
\end{equation}
and for dynamic time windows,
\begin{equation}
    r \times \sum_{m=m_{\rm cut}-1}^{\infty} e^{-rw} \dfrac{(rw)^m}{m!} \leq \textrm{FAR}.
    \label{eq:Fim_dyn}
\end{equation}
where the factor in front of the sum is the ``rate'' of time windows. For instance, for a background rate of $\SI{0.01}{\per\second}$ and a false alarm rate of $\SI{1}{\per\century}$, one gets multiplicity cuts of 7 and 8 events for the sliding and dynamic methods respectively (16 and 17 for $r=\SI{0.1}{\per\second}$).

Nevertheless, these methods are limited as neither take into account that signal bursts will have a very different time structure (localised in a few seconds interval) than the background (expected to be uniformly distributed within the time window).

Recently, there have been attempts in using additional time-related variables, such as the time difference $\Delta t$ between the first and the last event in a cluster~\cite{Casentini:2018bdf}: typically $\Delta t \sim \tau_{\rm phys}$ for signal-only cluster and $\Delta t \sim w$ for background-only cluster. This can however suffer from background contamination, in particular at high background rates.

In this publication, an alternative approach, that is getting rid of the traditional time window definition while benefiting from the time structure of signal burst, is proposed. The ~\autoref{sec:method} will present the new method. In~\autoref{sec:results}, the results when applied to toy SN simulation will be presented. In~\autoref{sec:nonpoisson}, it is shown that the method can still be applied in case of remaining non-Poisson background, with slight modifications. The \autoref{sec:discussion} presents a brief discussion about the different aspects of the new method.

\section{Method}
\label{sec:method}

Each selected neutrino event $i$ is characterised by a contribution $f(t; t_i)$, maximal at $t=t_i$. One can then compute: 
\begin{equation}
    \mathcal{F}(t) = \sum_i f(t; t_i)
\end{equation}
where the sum is performed over all events. This continuous function $\mathcal{F}$ can then be used to search for an excess i.e. $\mathcal{F}(t) > \mathcal{F}_{\rm cut}$. The value of the cut can be optimised to obtain a given false alarm rate. If $N$ centuries of background are simulated and the requested false alarm rate is one per century, one should find $\mathcal{F}_{\rm cut}$ such that $\mathcal{F}(t)$ exceeds this threshold less than $N$ times in total.

If one uses
\begin{equation}
f(t; t_i) = \left\{ \begin{array}{l}
     1 \text{ if } t_i - w < t < t_i \\
     0 \text{ otherwise}
\end{array} \right.,
\end{equation}
this is fully equivalent to the dynamic time window approach presented in \autoref{sec:intro}.

However, in this publication, the following function will be considered:
\begin{equation}
f(t; t_i) = \left\{ \begin{array}{l}
     e^{-(t-t_i)/T_c} \text{ if } t \geq t_i \\
     0 \text{ otherwise}
\end{array} \right.,
\end{equation}
where $T_c$ is a characteristic time. $\mathcal{F}(t)$ is re-written as:
\begin{equation}
    \mathcal{F}(t) = \sum_{i : t_i < t} \exp{\left(- \dfrac{t-t_i}{T_c} \right)}.
    \label{eq:RTS:nopenalty}
\end{equation}

For the implementation in an online process, one may simply store in memory the value $\mathcal{F}(t_{i})$ at the previous event and rewrite $\mathcal{F}$ for $t_i < t < t_{i+1}$:
\begin{equation}
    \mathcal{F}(t) = \exp{\left(- \dfrac{t-t_i}{T_c} \right)} \times \mathcal{F}(t_i),
\end{equation}
and:
\begin{equation}
    \mathcal{F}(t_{i+1}) = \exp{\left(- \dfrac{t_{i+1}-t_i}{T_c} \right)} \times \mathcal{F}(t_i) + 1.
\end{equation}

\emph{Average value:} In the case of Poisson-distributed background, the average value of $\mathcal{F}(t)$ is:
\begin{align}
    \langle \mathcal{F} \rangle(t) &= \int \mathcal{F}(t; t_1, \ldots, t_n) \prod_{i=1}^{n} P(t_i) dt_i \nonumber \\
    &= \sum_{i=1}^{n} \int e^{-(t-t_i)/T_c} \times P(t_i) dt_i,
\end{align}
where $n$ is the number of events within the period $[0, t]$ and $P(t_i)$ is the distribution of the time of event $i$. Using order statistic and assuming the events are uniformly distributed between $0$ and $t$, the latter may be written as:
\begin{equation}
    P(t_i) = \dfrac{n! \times (t_i/t)^{i-1} \times (1-t_i/t)^{n-i}}{t \times (i-1)! (n-i)!}.
\end{equation}

By taking the limit $t \to \infty$ ($n \to r t$), one can derive the full computation and obtain:
\begin{equation}
    \langle \mathcal{F} \rangle(t) \underset{t \to \infty}{\to} r T_c.
    \label{eq:meanF}
\end{equation}

This ensures that $\mathcal{F}$ remains stable in case of background-only events.

For simplicity, in the following, the method is named RTS$^2$ for \textit{Real-time Test Statistic for Supernovae}.

\begin{figure}[!ht]
    \centering
    \includegraphics[width=\linewidth]{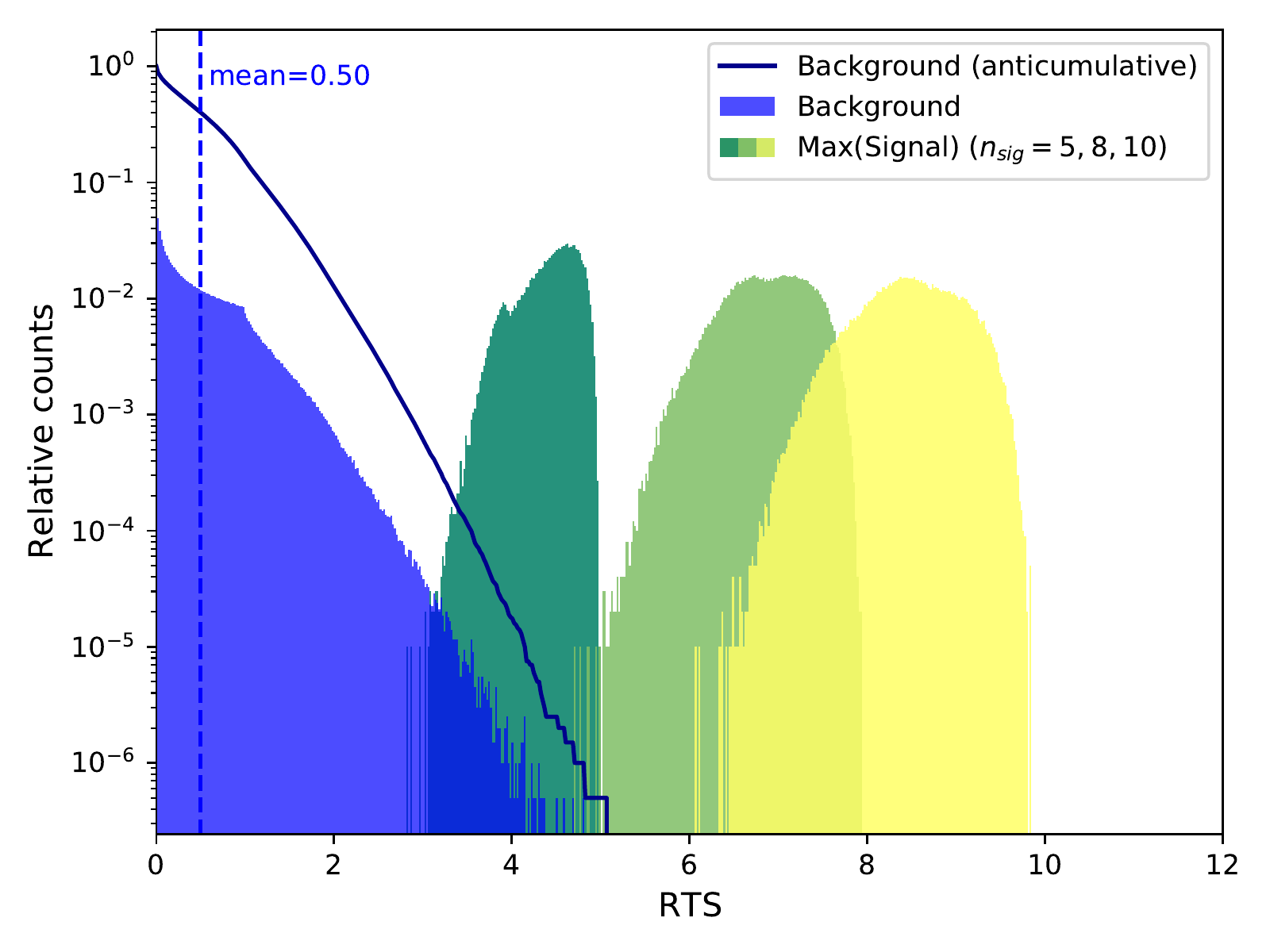}
    \caption{Distribution of $\mathcal{F}(t)$, computed every second on a background simulation with Poisson rate $r = \SI{0.05}{\per\second}$ and assuming $T_c = \SI{10}{\second}$, with its corresponding anticumulative (dark blue). The signal histograms (arbitrary scale) show the distributions of the maximal $\mathcal{F}$ values obtained for a signal injection of 5, 8 or 10 events ($\tau_1 = \SI{1}{\second}$, $\tau_2=\SI{0.01}{\second}$).}
    \label{fig:hist}
\end{figure}

The \autoref{fig:hist} shows the typical distribution of $\mathcal{F}(t)$ (using $T_c = \SI{10}{\second}$) if it is computed regularly for background-only scenario (with $r = \SI{0.05}{\per\second}$), compared with the distribution of the maximum value reached for a supernova signal (see \autoref{sec:signalsim} for the description of the latter). As expected from the \autoref{eq:meanF}, the background distribution keeps relatively small values and the average value is $\langle \mathcal{F} \rangle \sim 0.50 = r T_c$.

\section{Results}
\label{sec:results}

In this section, various background rates will be considered. The performance of the new method presented in \autoref{sec:method} will be compared to ones of the simple clusterisation in dynamic time windows (\autoref{eq:Fim_dyn}), referred as the ``multiplicity'' method.

In the following, $T_c = \SI{10}{\second}$ will be used as it is found to be a good compromise in the search for supernova signal with similar time scale.

\subsection{Signal simulation}
\label{sec:signalsim}

The simulation is performed in a generic way, so that it is relatively independent on the supernova models and on the neutrino detector:
\begin{itemize}
    \item A given number of signal events $n_S$ is supposed to be detected. This could be converted to a supernova distance for a given detector (fixed mass, target and efficiency) using SNOwGLoBES \cite{snowglobes} ;
    
    \item Their time distribution is parametrised by a double-exponential shape as in \cite{Casentini:2018bdf}:
    \begin{equation}
        p(t) = \left( 1-e^{-t/\tau_1} \right) \times e^{-t/\tau_2},
        \label{eq:signalpar}
    \end{equation}
    with typically $\tau_1 = \SIrange{10}{100}{\milli\second}$ and $\tau_2 \gtrsim \SI{1}{\second}$, representing respectively the rising time and the decaying time of the signal. In the following, when not explicitly specified, $\tau_1 = \SI{10}{\milli\second}$ and $\tau_2 = \SI{1}{\second}$ are used, which fit well the current SN models and SN1987a data.
\end{itemize}

Each signal burst is simulated on a short time interval (typically $\tau=\SIrange{30}{50}{\second}$), on top of simulated background events with rate $r_{\rm bkg}$. The signal efficiency for a given cut $\mathcal{C}$ is obtained by simulating a large number of signal bursts and by computing the fraction of bursts satisfying $\mathcal{C}$ during the period $\tau$ (one cluster passing the multiplicity cut or $\mathcal{F}(t)$ exceeding given threshold).

\subsection{Raw performance}

\begin{figure}[!ht]
    \centering
    \includegraphics[width=\linewidth]{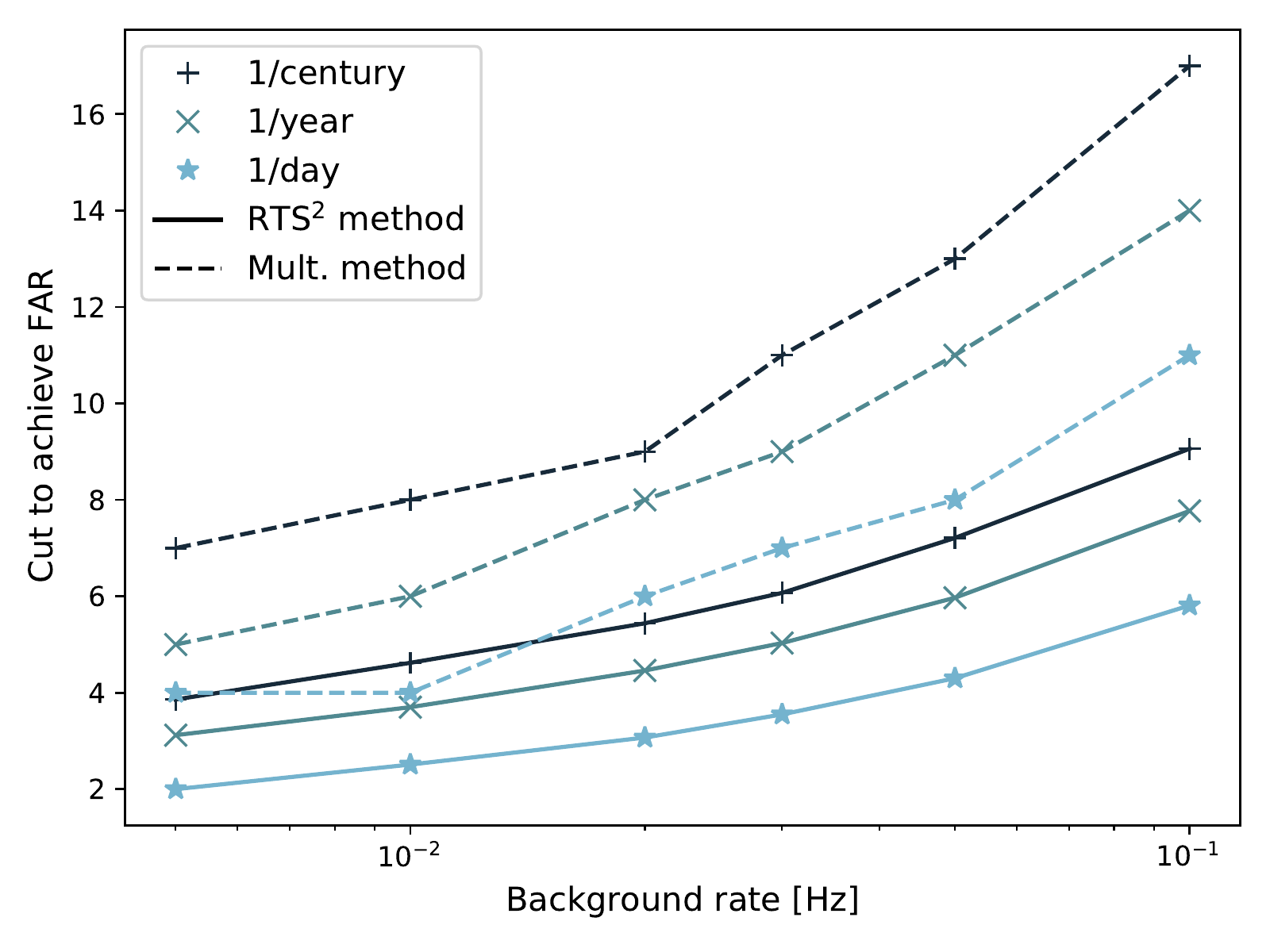}
    \caption{Evolution of the applied cuts on $\mathcal{F}$ for the RTS$^2$ method (solid line) and on multiplicity $m$ for the ``multiplicity'' method (dashed line) for different fixed false alarm rates (\SI{1}{\per\century}, \SI{1}{\per\year}, \SI{1}{\per\day}), as a function of fixed background rate.}
    \label{fig:res:cut_comp}
\end{figure}

Independently on the signal simulation, one can first compare directly the cuts applied on $\mathcal{F}$ and on multiplicity in order to achieve a given false alarm rate respectively in the RTS$^2$ and ``multiplicity'' methods, as illustrated in \autoref{fig:res:cut_comp}. This is relevant as these cuts correspond to the minimal number of events needed to trigger an alert.

The figure clearly shows that the new method is reaching lower multiplicities and is more robust with increasing multiplicities. The penultimate point on the right in \autoref{fig:res:cut_comp} corresponds to the scenario presented in the \autoref{fig:hist}, with $r=\SI{0.05}{\per\second}$ and with a typical cut $\mathcal{F} \gtrsim 4-5$ as already visible in the histogram.

\subsection{Signal selection efficiency}

\begin{figure}[!ht]
    \centering
    \includegraphics[width=\linewidth]{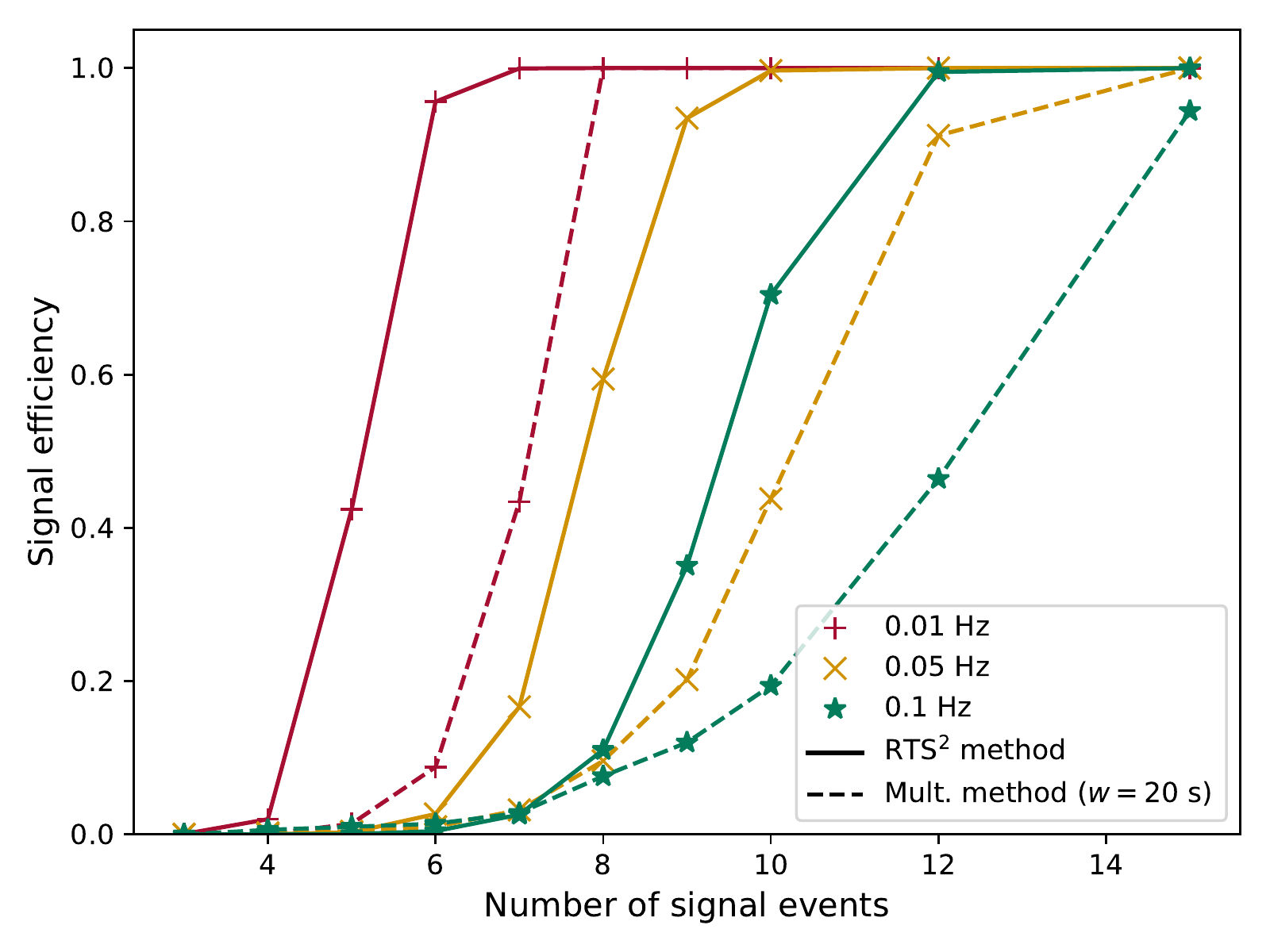}
    \caption{Comparison of the signal selection efficiency for the RTS$^2$ method (solid line) and for the ``multiplicity'' method (dashed line), as a function of the number of true signal neutrinos in the detected burst, for a fixed false alarm rate of \SI{1}{\per\century}, for different background rates (that impact both the simulation toys and the required cut).}
    \label{fig:res:sig_eff}
\end{figure}

The \autoref{fig:res:sig_eff} presents the obtained signal efficiency for different fixed background rates and number of injected signal neutrinos. It confirms the results presented in the previous paragraph: the RTS$^2$ method performs much better for low-multiplicity signal bursts and the effect is visible for all background rates. Referring back to the example of $r=\SI{0.05}{\per\second}$ from the \autoref{fig:hist}, one can achieve non-negligible efficiencies for $n_{\rm sig} > 6$, as seen in the clear separation between background and signal in the histogram.

For a fixed false alarm rate and background rate, one may then reduce the detection threshold (and therefore increase the distance horizon) of the supernova search.

\subsection{Stability with signal shape}

\begin{figure}[!ht]
    \centering
    \includegraphics[width=\linewidth]{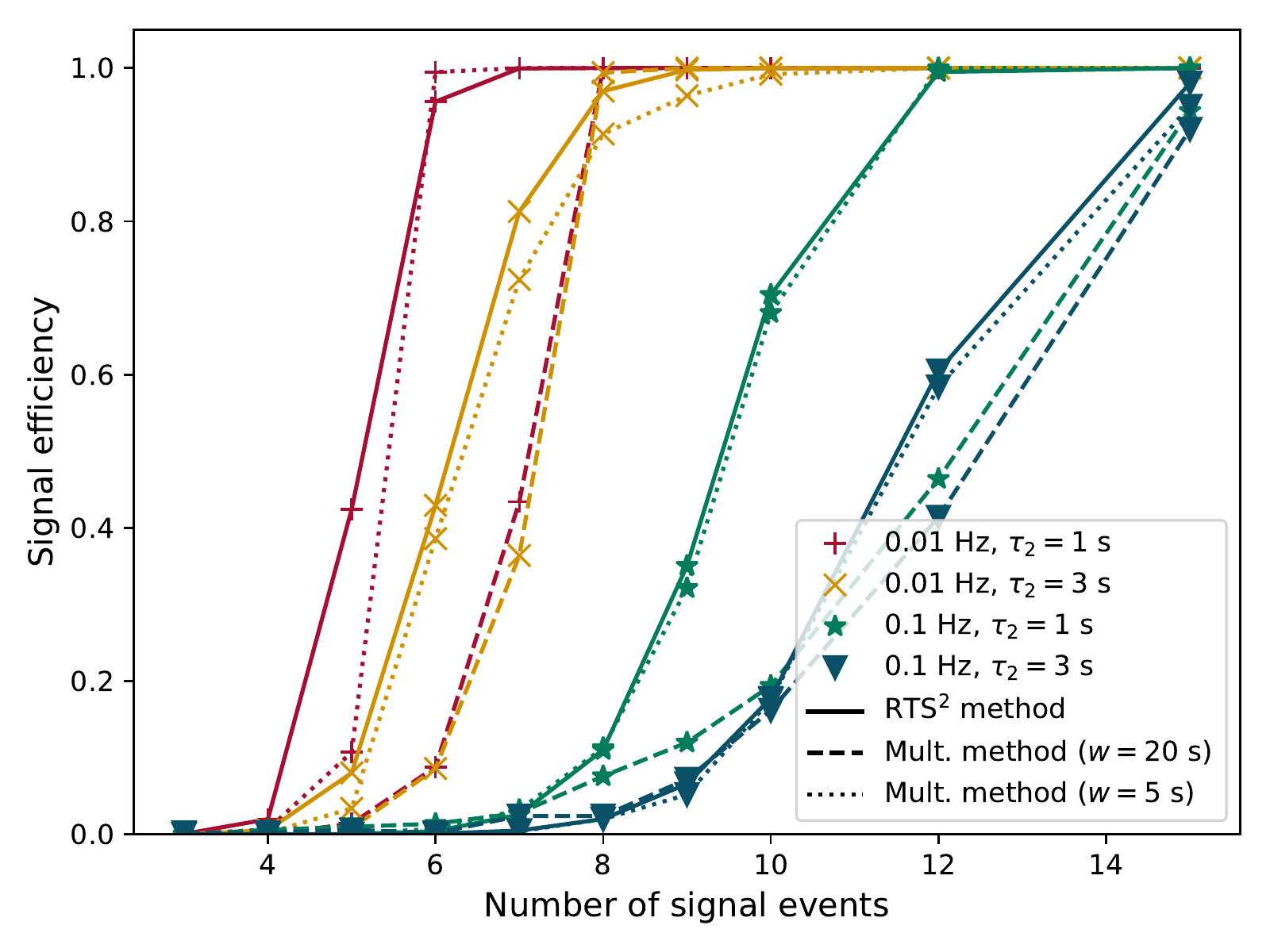}
    \caption{Comparison of the signal selection efficiency for the RTS$^2$ method (solid line) and for the ``multiplicity'' method (dashed line = time window of \SI{20}{\second}, dotted line = time window of \SI{5}{\second}), as a function of the number of true signal neutrinos in the detected burst, for a fixed false alarm rate of \SI{1}{\per\century}, different background rates $r = \SIrange{0.01}{0.1}{\per\second}$ and different values of the signal shape parameter $\tau_2$.}
    \label{fig:res:eff_dep}
\end{figure}

The \autoref{fig:res:eff_dep} illustrates how the signal efficiency changes with different signal shapes (\autoref{eq:signalpar}) and different sizes of time window for the ``multiplicity'' method. Reducing the time window down to \SI{5}{\second} allows increasing the efficiency with respect to the longer duration, but it does not reach the performance of the newly proposed method.

\section{Non-Poisson background}
\label{sec:nonpoisson}

Neutrino detectors may suffer from spatially correlated backgrounds whose times will not be distributed with a Poisson distribution. This is the case of spallation events, where radioactive decays may follow the interaction of high-energy cosmic ray muons with the matter of the detector. In Super-Kamiokande, this is handled by checking the spatial distribution of the identified vertices after clustering~\cite{Abe:2016waf}, computing the ``dimensionality'' of the vertex distribution and only selecting clusters identified as ``volume-distributed'' (vertices uniformly distributed in the volume of the detector).

One may try to adapt the real-time test statistic defined in \autoref{sec:method} in order to penalise events if they are detected nearby previous events, so that spatially-correlated clusters are effectively suppressed:
\begin{equation}
    \mathcal{F}_{SP}(t) = \sum_{i : t_i < t} \exp{\left(- \dfrac{t-t_i}{T_c} \right)} \times p_i,
    \label{eq:RTS:wpenalty}
\end{equation}
where $p_i$ is the weight applied to event $i$. A naive approach is to define:
\begin{equation}
    p_i = \displaystyle \prod_{j<i} \Big( 1 - e^{-\dfrac{t_i-t_j}{T_{\rm SP}}} e^{- \dfrac{\Vert \vec{x_i} - \vec{x_j} \Vert^2}{2\sigma_{\rm SP}^2}} \Big),
    \label{eq:spatial_penalty}
\end{equation}
where $\{ \vec{x_i} \}$ are event positions, $T_{\rm SP}$ and $\sigma_{SP}$ are the time and spatial characteristic of the spatial penalty and the sum runs over the events with $t_j < t_i$ (it may be sufficient to only consider events with $t_i - t_j \lesssim 10 T_{\rm SP}$). 

The formula ensures that $p_i \sim 1$ if there are no events nearby event $i$, while $p_i \to 0$ otherwise (event will not contribute to $\mathcal{F}$). The weight of a given event is reduced even further if more than one other event is detected nearby.

For a correlated cluster of $N$ events and characterised by a multi-normal distribution with $\sigma=\sigma_B$, one can derive that $\mathcal{F}_{SP}(t_c) \simeq \sum_{i=1}^N p_i \leq 2$, if $\sigma_B = \sigma_{\rm SP}$ and assuming all events are arriving at the same time $t_c$, even for $N \to \infty$. The effect of the cluster is effectively suppressed. The term $\sigma_{SP}$ can easily be tuned to the scale of the background of interest in a given detector.

\begin{figure}[!ht]
    \centering
    \includegraphics[width=\linewidth]{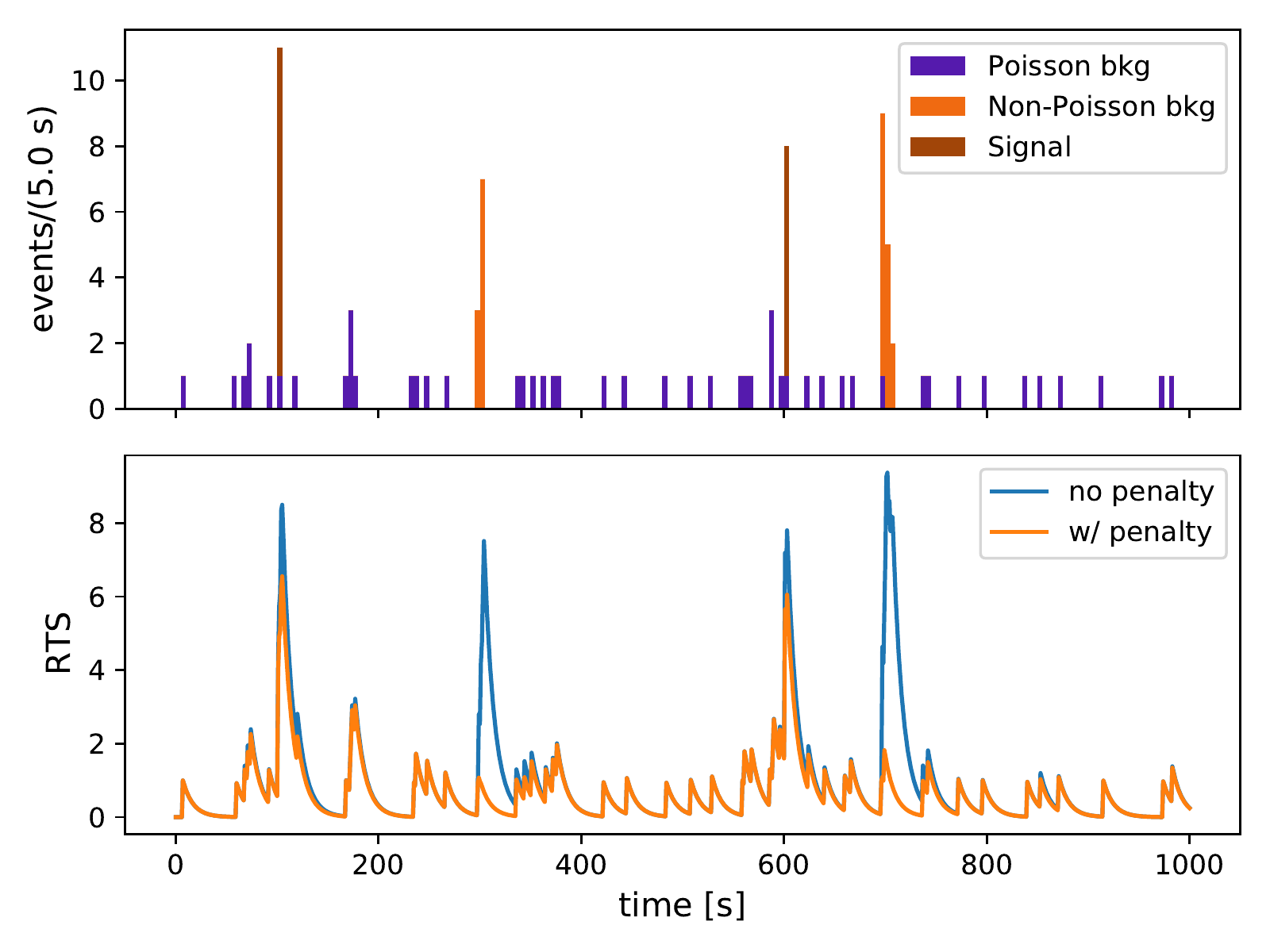}
    \caption{Illustration of non-Poisson rejection. The top panel shows an example of event rate as a function of time in a detector with the size of Super-Kamiokande. The injected Poisson background has a rate $r_{\rm bkg} = \SI{0.05}{\per\second}$. Two additional spatially-correlated clusters of background events (with $\sigma_B = \SI{2}{\meter}$) and with respectively $10$ and $15$ events) have been injected. Finally, two signal-like bursts with the parametrisation of \autoref{eq:signalpar} are added, with respectively 7 and 10 neutrinos. The bottom panel shows the test statistic time evolution for the RTS$^2$ method without and with the spatial penalty from \autoref{eq:spatial_penalty}.}
    \label{fig:np:illustrate}
\end{figure}

\begin{figure}[!ht]
    \centering
    \includegraphics[width=\linewidth]{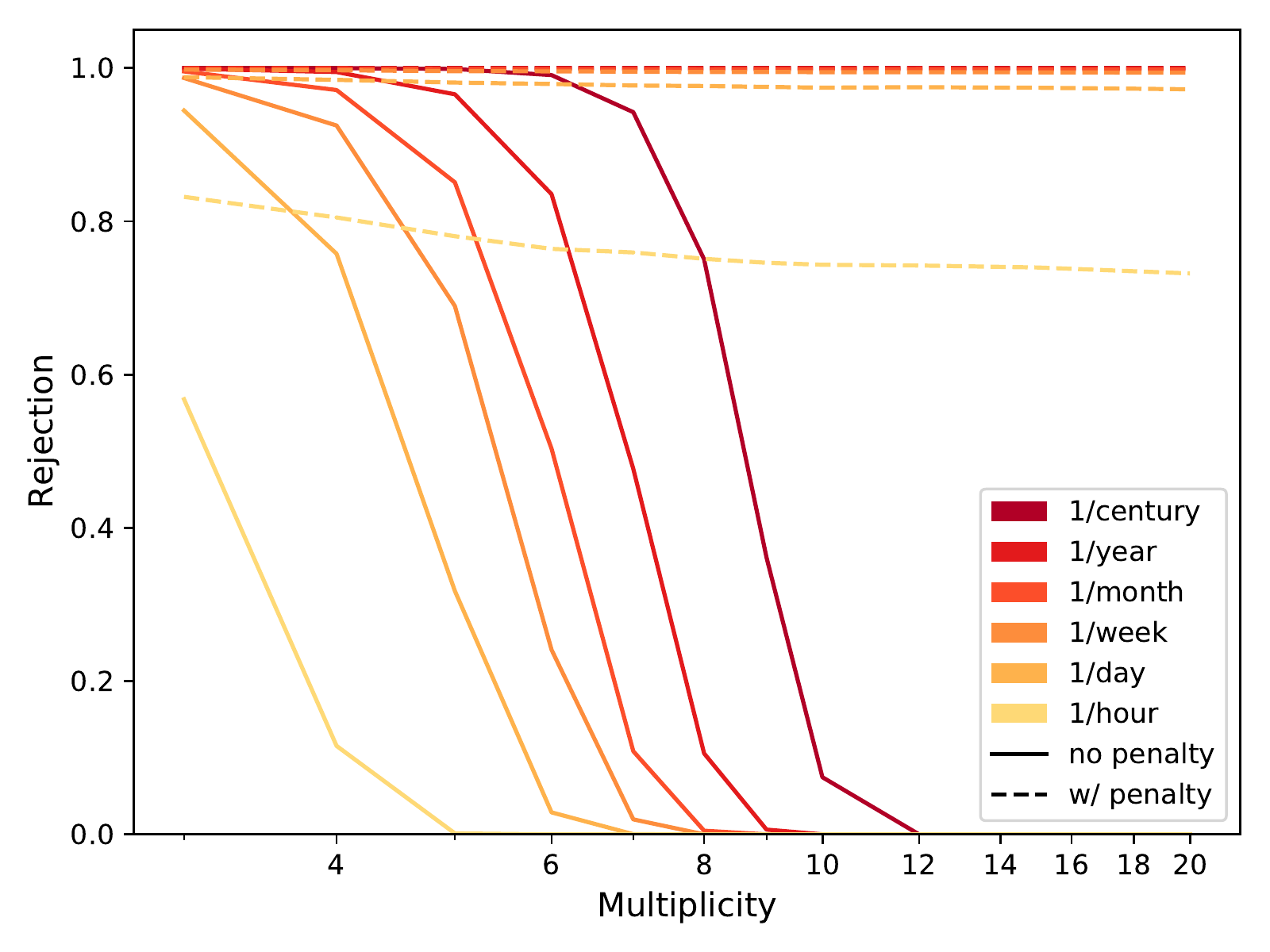}
    \caption{Fraction of simulated spatially-correlated clusters that are rejected by applying the cut on $\mathcal{F}(t)$ optimised in order to ensure a false alarm rate of 1/century, 1/year, 1/month, 1/week, 1/day or 1/hour from Poisson-distributed background (with $r_{\rm bkg} = \SI{0.05}{\per\second}$), as a function of injected multiplicity of the cluster. The solid lines (resp. dashed lines) show the results without (resp. with) the spatial penalty term in \autoref{eq:RTS:wpenalty}.}
    \label{fig:np:rej}
\end{figure}

The \autoref{fig:np:illustrate} illustrates how the implementation of the spatial penalty terms allows to reduce the contamination due to non-Poisson background, as peaks related to non-Poisson background (point-like sources) are shrunk, while signal bursts remain at the same level than without spatial penalty. The \autoref{fig:np:rej} presents the rejection power of RTS$^2$ method for spatially-correlated clusters.

One can see that, even though the chosen spatial penalty term is relatively simple, it is quite effective to reject point-source background. It is only in the case of low false alarm rates (e.g., 1/hour) that extra care may be needed. Furthermore, the definition of $p_i = p(\{x_j\}_{j \leq i}, \{t_j\}_{j \leq i})$ may be optimised to reject specific backgrounds related to a given experiment, as the proposed formula only takes care of point-like sources while e.g. spallation background may have vertices distributed along a line.

To classify each newly detected neutrino as part of a background ``burst'' ($p_i \to 0$) or not ($p_i \sim 1$), the use of a k-nearest neighbours algorithm or other outliers detection techniques may be very promising.

\section{Discussion}
\label{sec:discussion}

The presented approach would allow to select supernovae bursts with relatively low multiplicity ($\sim 5-10$), effectively increasing the horizon compared to more standard approaches. 

For reference, based on numbers from Table 4 of \cite{Kharusi:2020ovw}, the Hyper-Kamiokande detector would be able to reach M31-Andromeda ($d \sim \SI{778}{\kilo\parsec}$) while smaller current detectors would still trigger for most of the Milky Way satellite galaxies~\cite{Drlica-Wagner:2019vah} ($d \lesssim \SI{200}{\kilo\parsec}$).

The approach can be easily be implemented in a real-time monitoring system, as a recursive process of the type $\mathcal{F}(t_{i+1}) = a(t_{i+1}-t_i) \mathcal{F}(t_i) + p_{i+1}$, where $a(\tau)=e^{-\tau/T_c}$ characterises the decay of the signal and the $\{ p_{i} \}$ characterise the weights of individual events (either $=1$ or $<1$ if spatial penalties are applied). 

The threshold may be adjusted to safely reduce contamination by using Monte Carlo simulations of the background or by data-driven methods. The latter are particularly relevant if one wants to achieve false alarm rates of 1/(day, week, month), as limited datasets would be sufficient to validate it.

\section{Conclusion}
\label{sec:conclusion}

In this paper, the new RTS$^2$ method has been presented and applied to the detection of neutrino bursts with a generic time spectrum that fit well current supernovae models. The performances were compared to using solely a cut on the multiplicity in fixed size time window and it has been shown that the new method performs well, especially for higher background rates.

The method allows high flexibility on the choice of the function characterising the contribution of each event and it may be further improved by considering adding new variables in the analysis, such as the event energy.

It is possible to simply adapt the RTS$^2$ formula in order to take care of the non-Poisson distributed backgrounds, either with a basic penalty term (e.g. \autoref{eq:spatial_penalty}) or with specially adapted methods, to reject specific backgrounds in a given detector (outlier detection techniques or machine learning approaches).

Not only could it help to extend the horizon of existing and future large neutrino detectors beyond their current reach, but it may also be useful for smaller detectors that need a consistent way to define and send alerts to the community, in particular through SNEWS 2.0~\cite{Kharusi:2020ovw}. The latter aims to gather data from neutrino detectors all over the world (with various sizes and targets) for combination and initiate multi-messenger follow-up of a candidate supernova. This would ensure taking the most from the next galactic core-collapse supernovae that the astronomy community is waiting for.

\begin{acknowledgments}
I would like to thank the local group in the Department of Physics of the University of Padova, as well as the astrophysics working group of the Hyper-Kamiokande collaboration, for useful discussions. CloudVeneto is acknowledged for the use of computing and storage facilities. This project has received funding from the European Union's Horizon 2020 research and innovation programme under the Marie Skłodowska-Curie grant agreement No 754496.
\end{acknowledgments}

\bibliography{references}
\bibliographystyle{aasjournal}

\end{document}